\documentclass{elsarticle}

\usepackage{lineno,hyperref}
\modulolinenumbers[5]

\journal{Journal of \LaTeX\ Templates}

\newcommand{\bnabla}{\boldsymbol{\nabla}}

\newcommand{\bu}{\textbf{u}}
\newcommand{\bn}{\textbf{n}}

\newcommand{\ui}{\hat{\boldsymbol{\imath}}}
\newcommand{\uj}{\hat{\boldsymbol{\jmath}}}
\newcommand{\uk}{\hat{\textbf{k}}}

\newcommand{\pline}{$($------$)\hspace{0.1cm}$}

\usepackage{amsmath,amssymb}
\usepackage{subcaption}
\usepackage{relsize}
\usepackage{mathtools}
\usepackage{psfrag}

\begin{document}

\begin{frontmatter}

\title{A Second Order Thermal and Momentum Immersed Boundary Method for Conjugate Heat Transfer in a Cartesian Finite Volume Solver}


\author[label1]{Ryan Crocker }
\ead{rcrocker@uvm.edu}
\author[label1]{Yves Dubief}
\address[label1]{University of Vermont, Burlington, Vermont, USA}
\author[label2]{Olivier Desjardins}
\address[label2]{Cornell University, Ithaca, New York, USA}
\begin{abstract}
A conjugate heat transfer (CHT) immersed boundary (IB and CHTIB) method (IBM and CHTIBM) is developed for use with laminar and turbulent flows with low to moderate Reynolds numbers. The method is validated with the analytical solution of the flow between two co-annular rotating cylinders at $Re=50$ which shows second order error convergence of the $L_{2}$ and $L_{\infty}$ error norms, of the temperature field, over a wide rage of solid to fluid thermal conductivities, $\kappa_{s}/\kappa_{f} = \left(9-900\right)$. To evaluate the CHTIBM with turbulent flow a fully developed, heated, turbulent channel $\left(Re_{u_{\tau}}=150\text{ and } \kappa_{s}/\kappa_{f}=4 \right)$ is used which is correlated with to previous direct numerical simulation (DNS) results.  The CHTIBM is paired with a momentum IBM, both of which use a level set field to define the wetted boundaries of the fluid/solid interfaces and are applied to the flow solver implicitly with rescaling of the difference operators of the finite volume (FV) method (FVM).  
\end{abstract}

\end{frontmatter}

\section{Introduction} 

DNS has been a robust tool in numerical simulation and validation of new algorithms for fluid flow and heat transfer by comparing experimental and DNS results \cite{annrevmoin1999}.  Much attention has been paid to canonical flows, such as the DNS of fully developed turbulent channels in, \cite{kim1987tsf} and \cite{Mansour1999}, in these validation studies.  Fully developed flow and heat transfer have been studied experimentally in \cite{Khabakhpasheva1987}, \cite{Teitel1993} and \cite{sinai1987}, and with DNS \cite{Kawamura1998}, \cite{Kasagi1989}, \cite{Kasagi1992}, and \cite{kimmoin1989}, where numerical simulations were correlated with experimental data giving the proper scaling and simulation parameters for turbulent, heated, flows. 

The addition of CHT to DNS heat transfer studies greatly complicates the analysis of developed flows and initial research into the subject suffered from a lack of an adequate order of error.  Such as in, \cite{kang2009dbd} and \cite{kang2008}, where first order error convergence was observed.  Recently, advances in the interpolation schemes of curvilinear flow solvers has been shown to generate solutions with second order error convergence, \cite{Nagendra2014}.  But this was shown for small ratios of thermal conductivity and was not applied to the solution implicitly or in a way easily applicable to cartesian FV solvers with complex, non-mesh conforming boundary conditions.

Non-mesh conforming boundary conditions, IBMs, have been used to simulate complex geometries with moderate Reynold's number flows since their development by Peskin,\cite{peskin1972}, to simulate heart valves.  Yusof expanded the idea of IBs and developed the direct forcing method, \cite{Yusof:1997}, eliminating velocity stability issues associated with non-direct forcing IBMs.  The final issues with stability linked to small cut cells and pressure oscillations \cite{Lee2011} were solved with by the reconstruction or cut-cell IBMs where multiple methods were developed: cell merging \cite{Ye1999}, mixed cell linking/merging \cite{mittal2008versatile},  cell linking\cite{Kirkpatrick2003}, and cell mixing \cite{Meyer2010} methods, all of which show second order error in their results for low to moderate Reynold's numbers. In this study the sharp representation of boundaries is done through a mixing method modified from \cite{Meyer2010}.

In the present work a DNS flow solver, developed in \cite{Desjardins2007}, with CHTIB and momentum IB algorithms, with scalar transport, is used to study the effect of CHT on two separate canonical flows. With the domains that are separated with level set fields from which the cut-cell geometric data is derived.  The first being a heated co-annular cylinder set up in two dimensions with a simple analytical solution which shows second order accuracy for very large solid/fluid conductivity ratios at low Reynolds numbers.  The second half of this study uses a fully developed channel heated from the top and the bottom boundaries where the root-mean-square of the near-boundary fluctuating temperature is shown to correlate well with previous results from \cite{Tiselj2001}. 

\section{Governing Equations} 
The basic equations governing the transport of momentum and the scalar (temperature) carried by the flow are given in their most general forms in sections \ref{sec:Motrans} and \ref{sec:SCtrans}.  The are kept fully conservative and will subsequently be non-dimensionalized for each numerical experiment.

\subsection{Momentum and mass conservation}\label{sec:Motrans}
The two equations used to simulate fluid flow are the Navier-Stokes set of equations, 
\begin{equation}
\partial_{t}\left(\rho\textbf{u}\right) + \bnabla \cdot \rho \left( \textbf{u} \otimes \textbf{u} \right)= -\bnabla \cdot \textbf{I}P+ \bnabla \cdot \left( \mu \bnabla \textbf{u}\right), \label{eq:navier_stokes_dim}
\end{equation}
and the conservation of mass, or continuity equation,
\begin{equation}
\partial_{t}\left(\rho\textbf{u}\right) +\bnabla \cdot \left( \rho\textbf{u}\right)=0.\label{eq:continuity}
\end{equation}
In equations  \ref{eq:continuity} and \ref{eq:navier_stokes_dim}, bold terms are vectors, $\textbf{u}$ is the velocity vector $\left(\textbf{u} =u \ui+v \uj+w\uk \right)$, $P$ is the pressure, $\mu$ is the dynamic viscosity, $\rho$ is the density, $\textbf{I}$ is identity matrix, and $\partial_{t}$ is the temporal derivative.

\subsection{Scalar transport}\label{sec:SCtrans}
Temperature is transported as a passive scaler in the flow, via equation \ref{eq:energy},
\begin{equation}
\partial_{t}\phi+\bnabla \cdot \left(\textbf{u}\phi\right) = \bnabla \cdot \left( \alpha_{f} \bnabla \phi\right),  \label{eq:energy}
\end{equation}
where $\phi$ is the temperature, and $\alpha$ is the thermal diffusivity.  For heat transfer inside the solid domain the convective term is dropped from equation \ref{eq:energy} giving, 
\begin{equation}
\partial_{t}\phi = \bnabla \cdot \left( \alpha_{s} \bnabla \phi\right),  \label{eq:energy_solid}
\end{equation}
where the $f$ and $s$ subscripts stand for the fluid and solid domains, respectively. 

\section{Sub-Domain by Level Set Field}
The fluid and solid domains are separated by different signed regions of a level set field \cite{OshSeth1988}, $G$, over the zero isosurface, $\Gamma \rightarrow G=0$.  The use of a particular form of the level set field, a signed distance function, allows the immersed boundaries to represent a non-mesh conforming boundary on a cartesian grid.  The main property of a signed distance function, $\bnabla G|=1$, allows this by giving a definite a zero iso-surface, $G=0\rightarrow\Gamma$, where at all times a normal is defined by $\textbf{n}=\nabla G/|\nabla G|$, and easily accessible geometric information for cut-cell areas (see figure \ref{fig:cut_cell_HT}) and volumes needed in section \ref{sec:num_meth}.  The geometric information is found with a simplex marching algorithm which uses the intersection of tetrahedra with $\Gamma$.  This information is then used to locate the cut cell intersections with $\Gamma$ and the resulting cut-cell geometry \cite{Desjardins_spray2013}.  The last caveat of level set domain definition is that sign of the $G$ also sets the parameters of each subdomain, where $G\ge0$ is the solid subdomain and $G<0$ is the fluid domain, but the choice of this is arbitrary. 
\section{Numerical Methods}\label{sec:num_meth}
\begin{figure}
\begin{subfigure}{.5\textwidth}
  \centering
  \includegraphics[width=1.0\linewidth, clip=true]{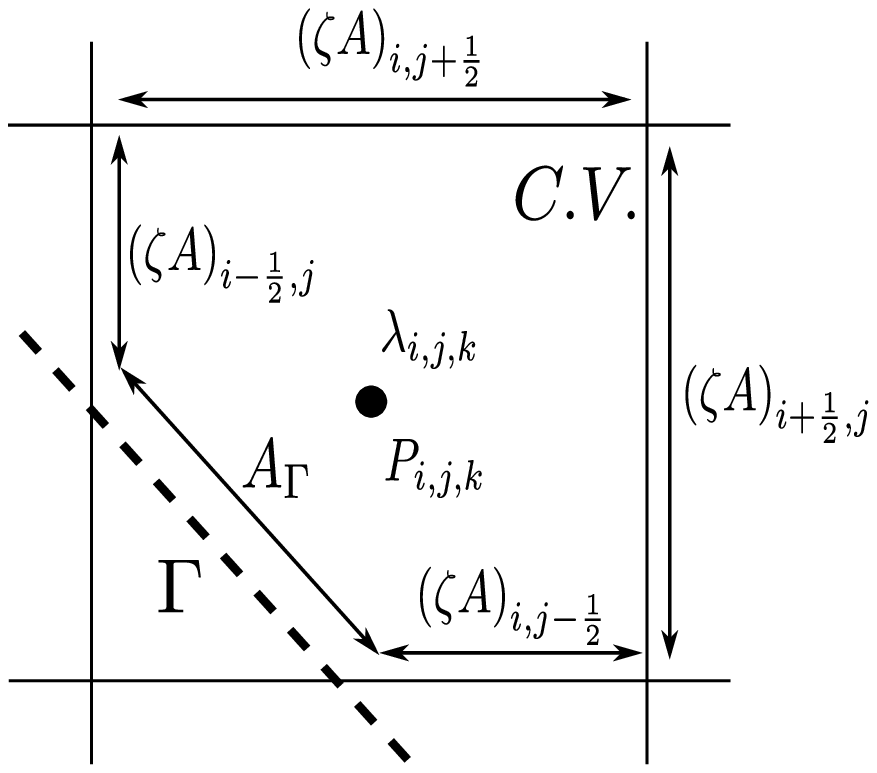}
  \caption{}
  \label{fig:cut_cell_moment}
\end{subfigure}
\begin{subfigure}{.5\textwidth}
  \centering
  \includegraphics[width=0.9\linewidth, clip=true]{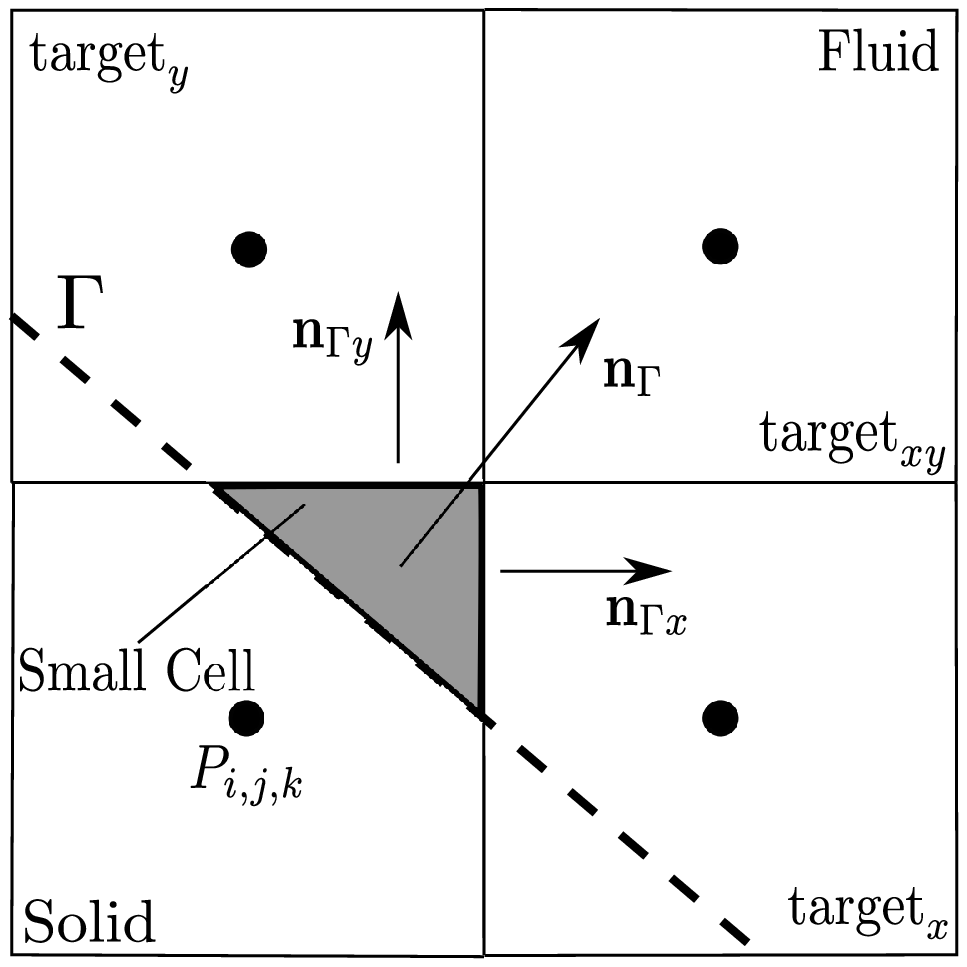}
  \caption{}
  \hfill
  \label{fig:cut_cell_small}
\end{subfigure}
\caption{a.) A schematic of a 2D cut cell with the IB surface shown. b.) A schematic of the small cell mixing used to fix the cells that do not fall under the $\epsilon$ criteria.}
\label{fig:cut_cell_IB_mix}
\end{figure}

The general conservation form of a transport equation is a typical first step in the derivation of the finite volume method.  As the IB methods for both energy and momentum are themselves derived from the finite volume method equations \ref{eq:continuity} and \ref{eq:energy} are rearranged into general conservation equations on a staggered pressure/velocity grid (see figures \ref{fig:cut_cell_moment} and \ref{fig:cut_cell_noHT}).   This begins with definition of the flux function, $F$, for momentum the flux function is a tensor, 
\begin{equation}
[\textbf{F}\left(P,\textbf{u}\right) ] =  \left(\rho \textbf{u}  \otimes \textbf{u} \right) + \textbf{I}P-\left( \mu \bnabla \textbf{u}\right), \label{eq:navier_stokes_flux}
\end{equation}
and for temperature transport it is a vector,
\begin{equation}
\textbf{F}\left(\phi,\textbf{u}\right) =  \left( \alpha \bnabla \phi\right)-\left(\textbf{u}\phi\right),  \label{eq:energy_flux}
\end{equation}
Equations \ref{eq:navier_stokes_flux} and \ref{eq:energy_flux} are now in the form of general conservation equations, 
\begin{equation}
\underbrace{\partial_{t}\Theta+\bnabla \cdot \textbf{F} = 0}_{energy} \hspace{0.75cm} \text{or} \hspace{0.75cm} \underbrace{\partial_{t}\boldsymbol{\Theta}+\bnabla \cdot [\textbf{F}] = 0}_{momentum} \label{eq:gen_mo_en}
\end{equation}
The equations in \ref{eq:gen_mo_en} are also known as the advective form of the linear conservation law, balance between the diffusive transport budget and the convective transport budget, for time $t$, gives the overall change of the conserved quantity, $\Theta$, as a function of time.

\subsection{Basic finite volume discretization}\label{sec:fv_basic}
Using the fully conservative form of the momentum equations gives a second order accurate discretization of equation \ref{eq:navier_stokes_dim} , as shown in \cite{Morinishi1998}, \cite{Morinishi2004} and \cite{Desjardins2007}, for low mach number flows.  Discretizing equation \ref{eq:energy} in the same manner as that shown in \cite{osher1982} will also give second order accuracy, with high-ordered-upwinded-centered scheme (HOUC), \cite{henrick2005}.  The basic finite volume (FV) method (FVM) can most generally be derived with the momentum form of the conservation equation given in equation \ref{eq:gen_mo_en}. 

\begin{figure}
\begin{subfigure}{.5\textwidth}
  \centering
  \includegraphics[width=1.0\linewidth, clip=true]{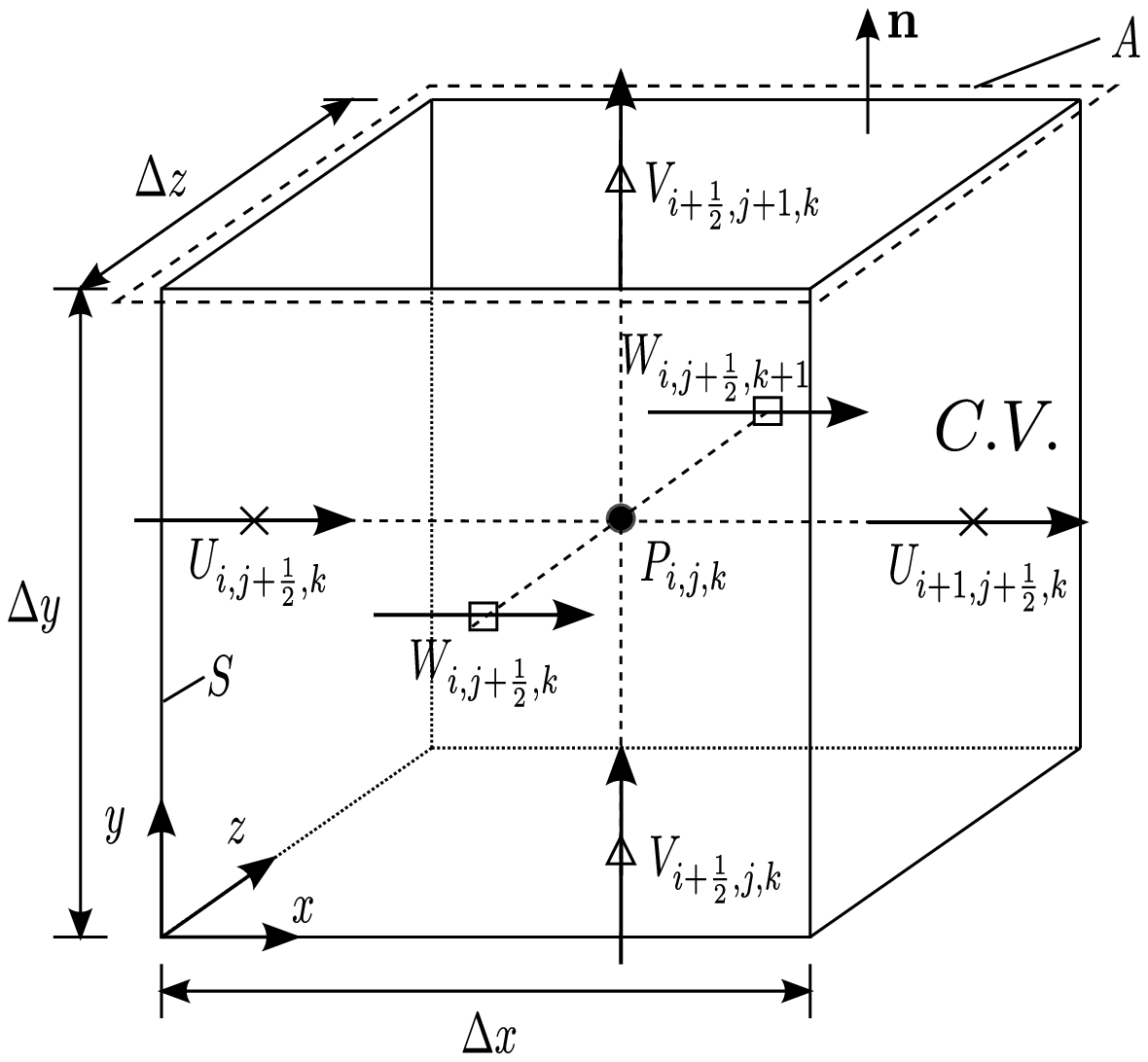}
  \caption{}
  \label{fig:cut_cell_noHT}
\end{subfigure}
\begin{subfigure}{.5\textwidth}
  \centering
  \includegraphics[width=1.0\linewidth, clip=true]{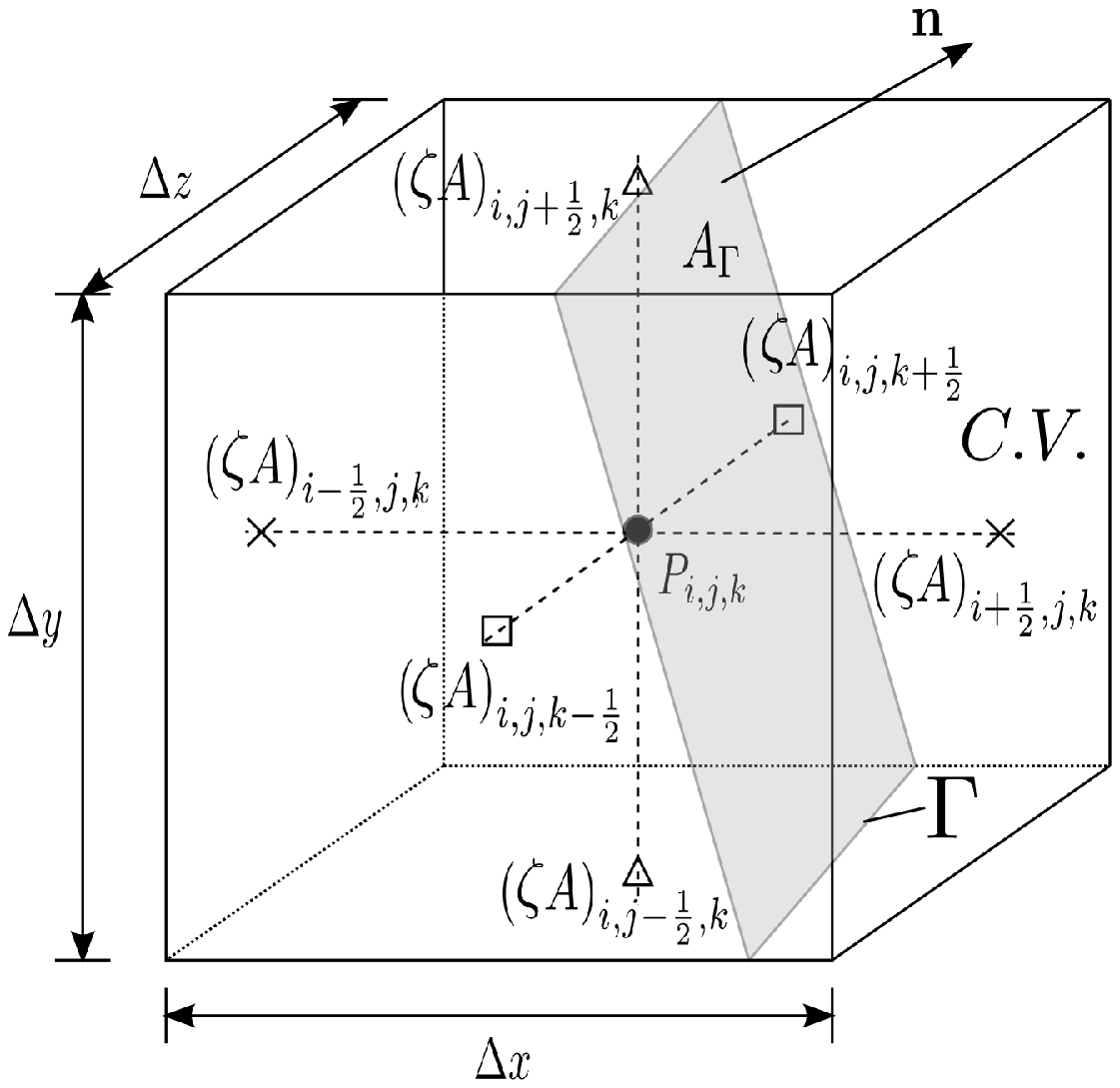}
  \caption{}
  \label{fig:cut_cell_HT}
\end{subfigure}
\caption{a.) A computational, finite volume, cell for momentum, with temperature and pressure carried at the center node.  b.) A 3D cut cell for the thermal or momentum IB showing a more general 3D cut cell.}
\label{fig:cut_cell_IBHT}
\end{figure}

Integrating equation \ref{eq:gen_mo_en} over the C.V. in figure \ref{fig:cut_cell_noHT} and replacing the conserved quantity, $\boldsymbol{\Theta}$ with $\textbf{u}$ and dropping the tensor brackets around $F$, gives the integral form of the conservation law for momentum,
\begin{equation}
\int_{V}\partial_{t}\left(\rho\bu\right) dV+ \int_{V}\bnabla \cdot \textbf{F} dV=0 \hspace{0.25cm} \rightarrow \hspace{0.25cm}  \partial_{t}\overline{\bu} = \frac{1}{V}\oint_{S}  \textbf{F} \cdot \bn dA, \label{eq:con_int_gen_V}
\end{equation}
where $V=\Delta x\Delta y \Delta z$, $\overline{\bu}=\int_{V} \bu dV$ is the average velocity over the volume, $A$ is the side area and $\bn$ the cell face normal from the C.V. in figure \ref{fig:cut_cell_noHT}, and Gauss's law has been used from left to right.  Omitting the over bar on $\textbf{u}$, and integrating around the surface $S$, the expanded, spatially discretized form of equation \ref{eq:con_int_gen_V} is, 
\begin{equation}
\begin{split}
\partial_{t} \left(\rho\bu\right)= 
              \frac{1}{V}   \bigg[ &\left(\textbf{Fx}_{i+\frac{1}{2},j,k}A_{i+\frac{1}{2},j,k}-\textbf{Fx}_{i-\frac{1}{2},j,k}A_{i-\frac{1}{2},j,k}\right)\cdot \textbf{n}_{x} \\
                                            +&\left(\textbf{Fy}_{i,j+\frac{1}{2},k}A_{i,j+\frac{1}{2},k}-\textbf{Fy}_{i,j-\frac{1}{2},k}A_{i,j-\frac{1}{2},k}\right) \cdot \textbf{n}_{y}  \\
                                            +&\left(\textbf{Fz}_{i,j,k+\frac{1}{2}}A_{i,j,k+\frac{1}{2}}-\textbf{Fz}_{i,j,k-\frac{1}{2}}A_{i,j,k-\frac{1}{2}}\right) \cdot \textbf{n}_{z} \bigg] , \label{eq:FV_noib}
\end{split}
\end{equation}
where $A$ is the area of the cell face, $A_{i,j \pm\frac{1}{2},k}=\Delta x\Delta z$, $A_{i\pm\frac{1}{2},j,k}=\Delta y\Delta z$, $A_{i,j,k\pm\frac{1}{2}}=\Delta x\Delta y$, $\bn=[\textbf{n}_{x}\hspace{0.25cm}\textbf{n}_{y}\hspace{0.25cm}\textbf{n}_{z}]$ are the normal components from each face,  and $\textbf{F}=\left[\textbf{Fx} \hspace{0.25cm}\textbf{Fy}\hspace{0.25cm}\textbf{Fy}\right]$ are the flux functions components in the $x$, $y$, and $z$ $\left(\ui \hspace{0.25cm} \uj \hspace{0.25cm} \uk \right)$ directions. 

Replacing the flux function and the time derivative in equation \ref{eq:FV_noib} with the flux function from equation \ref{eq:energy_flux} and the time derivative from equation \ref{eq:energy}, respectively,  gives the FV formulation of the energy transport equation. 

\subsection{Conservative Temporal Discretization}\label{sec:time_disc}
Time advancement of equation \ref{eq:FV_noib} is done by use of a 3 directional semi-implicit fractional step method, \cite{Kim1985}, using a second order in time Crank-Nicholson discretization scheme, with approximate factorization, \cite{moin1994}, which is a form of alternating direction implicit methods (ADIM).  The ADIM system of equations is trilinear and solution of which is readily expanded to three dimensions, and to codes written in parallel with message passing interfaces (MPI), 
\begin{equation}
\partial \left(\rho \bu \right) = \frac{\left(\rho \bu \right)^{n+1}- \left(\rho \bu \right)^{n}}{\Delta t}=\textbf{L}\left(\bu,P\right) ,\label{eq:con_time}
\end{equation}
where $n$ denotes the present time step, $n+1=t+\Delta t$, and $\textbf{L}$ is the linear operator $\textbf{L}\left(\bu,P\right) =-\bnabla \cdot \textbf{F}\left[\frac{1}{2}\left(\bu^{n+1}+\bu^{n}\right),P^{n+1}\right]$.  Newton-Iterations, $m$, are applied to the sub-steps of equation \ref{eq:con_time}, for each time step, $\Delta t$.  The method is taken from \cite{pierce2001} and is used here, as an example in the $x$ direction, as, 
\begin{equation}
\begin{split}
&\left[ \textbf{I} -\frac{1}{2}\Delta t \partial_{\bu}\textbf{L}\left(\bu,P\right) \right]\left( \rho u^{n+1}_{m+1} -\rho u^{n}_{m} \right) =\\
&\rho u^{n}_{m=0} -\rho u^{n}_{m}+\Delta t \textbf{L}\left[\frac{1}{2}\left(u^{n}+u^{n+1}\right),Px^{n+1}_{m}\right],\label{eq:con_time_two}
\end{split}
\end{equation}
where $m=0$ is the initial solution at the present time step, $n$. 

The velocity field resulting from the solution of equations \ref{eq:con_time}-\ref{eq:con_time_two}, using equation 10 \ref{eq:FV_noib}, will not be solenoidal.  To get a divergence free velocity equation \ref{eq:continuity} is used to project the pressure onto a divergence free velocity field give the Pressure-Poisson equation,   
\begin{equation}
\nabla^{2} \eta = -\bnabla  \cdot \bu^{*} \rightarrow \bu^{n+1}= \bu^{n}-\nabla\eta \hspace{0.25cm}\text{and}\hspace{0.25cm} P^{n+1}=P^{n}+\frac{\eta}{\Delta t} \label{eq:PP}
\end{equation}
where $u*=u^{n+1}$ is the intermediate velocity from the fractional step method, $P^{n+1}$ is the updated pressure,  and $\eta$ is the pseudo-pressure field also from the fractional step method. 

The time advancement scheme given in equations \ref{eq:con_time} and \ref{eq:con_time_two} is directly applicable to the scalar transport spatial discretization given given in sections \ref{sec:num_meth} and \ref{sec:fv_basic}.  With the only difference being the replacement of the flux function in the linear operator $\textbf{L}$ with $\textbf{L}\left(\phi,\bu\right)=-\bnabla \cdot \textbf{F}\left(\phi,\textbf{u}\right)$, with the flux function from equation \ref{eq:energy_flux}, and the fractional step method is not used, as the continuity equation is not applied. 

\subsection{Momentum Immersed Boundary Method}
The second order accurate, momentum cut-cell IB will be covered, briefly, in this section.  Much more in depth coverage of the topic can be found in \cite{Meyer2010}, \cite{Desjardins_spray2013} and \cite{Zhen2014}.  Nominally, the IB methods for momentum and thermal transport provide the same functionality.  They add cells cut by a non-grid-conforming boundary condition back into the time evolution solutions, minimizing error.  Where they differ in that the thermal IB from section \ref{sec:therm_IB} requires the information on both sides of the surface $\Gamma$, where the momentum cut-cell IB, given here, is only concerned with the wetted side of $\Gamma$. 

Cutting the control volume cells adds two new terms to the momentum equations, the viscous, $D$, and momentum, $C$, source terms. $D$ and $C$ enforce the no-slip condition tangental to  $\Gamma$, and the Lagrangian velocity of $\Gamma$,  respectively.   To fully describe the FVM with cut cells, along with $D$ and $C$, the cut cell aperture size, $\left(\zeta A\right)_{i,j,k}$, the area fraction $\zeta_{i,j,k}$, and the cut cell volume fraction, $\lambda_{i,j,k}$, and $V_{i,j,k}=\lambda_{i,j,k} \Delta x \Delta y \Delta z$ is the cell volume.   The addition of the IB will add the two new terms to equation \ref{eq:FV_noib}, 
\begin{equation}
\begin{split}
\partial_{t}\left(\rho\bu\right) = &+\frac{1}{\lambda_{i,j,k}\Delta x}\left[\left(\zeta A\right)_{i+\frac{1}{2},j,k}\textbf{Fx}^{n}_{i+\frac{1}{2},j,k}-\left(\zeta A\right)_{i-\frac{1}{2},j,k}\textbf{Fx}^{n}_{i-\frac{1}{2},j,k}\right] \cdot \textbf{n}_{x} \\
                                                &+\frac{1}{\lambda_{i,j,k} \Delta y}\left[\left(\zeta A\right)_{i,j+\frac{1}{2},k}\textbf{Fy}^{n}_{i,j+\frac{1}{2},k}-\left(\zeta A\right)_{i,j-\frac{1}{2},k}\textbf{Fy}^{n}_{i,j-\frac{1}{2},k}\right]\cdot \textbf{n}_{y} \\
                                                &+\frac{1}{\lambda_{i,j,k} \Delta z}\left[\left(\zeta A\right)_{i,j,k+\frac{1}{2}}\textbf{Fz}^{n}_{i,j,k+\frac{1}{2}}-\left(\zeta A\right)_{i,j,k-\frac{1}{2}}\textbf{Fz}^{n}_{i,j,k-\frac{1}{2}}\right]\cdot \textbf{n}_{y} \\
                                                &+\frac{1}{\lambda_{i,j,k}V_{i,j,k}}\left[C+D\right]. \label{eq:ib_mom}
\end{split}
\end{equation}
Both $\zeta_{i,j,k}$ and $\lambda_{i,j,k}$ are $0 < \zeta_{i,j,k} \text{ , }\lambda_{i,j,k} \le 1$ and are the fraction of the cell face and volume lost due to the cell cut (see figures \ref{fig:cut_cell_moment} and \ref{fig:cut_cell_HT}).

The viscous term added by the IB method, $D$, comes from the integration of the equation \ref{eq:con_int_gen_V} taking into account the interface $\Gamma$  from figure \ref{fig:cut_cell_moment}, giving, 
\begin{equation}
\ \int_{\Gamma}= \nu D= \nu \left(\nabla\textbf{u}\right) \cdot \textbf{n}_{\Gamma} dA_{\Gamma}. \label{eq:IB_visc}
\end{equation}
Equation \ref{eq:IB_visc} accounts for the no slip boundary condition at the surface cut by the boundary.  The momentum exchange term, $C$, accounts for any motion of the interface cut.  If the boundary is moving $C$ has the form, 
\begin{equation}
\ C=v_{\textbf{n}} \frac{V_{i,j,k}}{\Delta t}. \label{eq:IB_mome}
\end{equation}
If the interface boundary is not moving and impervious, $C=0$.  When the boundary is in motion the wall normal velocity, $v_{\textbf{n}}$ is non-zero. 

The application of the cut cell method can lead to very small cell volumes compared to the rest of the mesh.  By not taking account of these small cells the velocity and pressure solutions will be numerically unstable without, small, and computationally prohibitive,  time steps.  The most efficient and stable numerical treatment of the small cells is by merging, \cite{Mittal1999ib},  them with the surrounding cut cells and un-cut cells.  Though giving a numerically stable solution cell merging does not lend itself to straight forward implementation with IB's that are not stationary.  A more versatile method is to mix the small cell values with the cells that surround them such as the method used in \cite{Adams2006}.  The mixing method is altered to fit a staggered grid and again follows the method in \cite{Meyer2010}.

The mixing fraction, $\beta^{i}_{i,j,k}$ and the exchange variable, $\xi_{i}$, are given in the $x$ direction with the conservative mixing variable, $\Upsilon$, in equation \ref{eq:con_mix}
\begin{equation}
\ \xi_{x}=\frac{\beta^{x}_{i,j,k}}{\beta^{x}_{i,j,k}V_{i,j,k} +V_{\text{target}_{x}} } \left(V_{i,j,k}\Upsilon^{*}_{\text{target}_{x}}-V_{\text{target}_{x}}\Upsilon^{*}_{\text{target}_{x}}\right). \label{eq:con_mix}
\end{equation}
Mixing fractions, $\beta^{M}_{i,j,k}$, are a combination of the normal vector to the target cell and $\lambda_{i,j,k}$, which in two dimensions is, 
\begin{equation}
\begin{split}
 &\beta^{M}_{i,j,k} = \textbf{n}^2_{M}\lambda_{i,j,k} \rightarrow \\
 &\beta^{x}_{i,j,k} =  \textbf{n}^2_{\Gamma x}\lambda^{\text{target}_{\text{\scalebox{1.5}{$x$}} }}_{i,j,k} \hspace{0.5cm} \beta^{y}_{i,j,k} =\textbf{n}^2_{\Gamma y}\lambda^{\text{target}_{\text{\scalebox{1.5}{$y$}} }}_{i,j,k} \hspace{0.5cm} \beta^{xy}_{i,j,k} =\textbf{n}_{\Gamma x}\textbf{n}_{\Gamma y}\lambda^{\text{target}_{\text{\scalebox{1.5}{$xy$}} }}_{i,j,k} .\label{eq:beta_mix}
\end{split}
\end{equation}
The cut off for cells to be considered small for this work is $\lambda_{i,j,k} \le 0.1$, and the condition that the sum of all of the mixing fractions be normalized so that, $\sum^{N}_{M=1}\beta^{M}_{i,j,k}=1$, where $N$ is the number of target cells.  The values for $\Upsilon$ are mixed to the surrounding cells from the small cells by equation \ref{eq:con_mix_sum},
\begin{equation}
\ \Upsilon = \Upsilon^{*} +\frac{1}{V_{i,j,k}} \sum^{N}_{M=1}\xi_{M}, \label{eq:con_mix_sum}
\end{equation}
where in this case $\Upsilon^{*}$ is the variable to be mixed to before mixing, and $N$ is the number of surrounding cells to the small cell. The form of $\Upsilon^{*}$ that is gives the best solution to the pressure is, 
\begin{equation}
\ \Upsilon^{*} =\frac{1}{V_{i,j,k}} \int_{V_{i,j,k}}\textbf{F}\cdot \textbf{n}, \label{eq:con_mix_int}
\end{equation}
and will also be the method in which the small cell conservative fluxes are mixed to their neighbors.  After mixing, the fully discretized Pressure-Poisson equation (equation \ref{eq:PP}), from the fractional step method, is written as equation \ref{eq:disc_poiss}, 
\begin{equation}
\begin{split}
              &\frac{1}{\Delta x}\left[\left(\zeta A \right)_{i+\frac{1}{2},j}\frac{\partial \eta}{\partial x}\Bigg\vert_{i+\frac{1}{2},j}-\left(\zeta A \right)_{i-\frac{1}{2},j}\frac{\partial \eta}{\partial x}\Bigg\vert_{i-\frac{1}{2},j} \right]  \\
            +&\frac{1}{ \Delta y}\left[\left(\zeta A \right)_{i,j+\frac{1}{2}}\frac{\partial \eta}{\partial y}\Bigg\vert_{i,j+\frac{1}{2}}-\left(\zeta A \right)_{i,j-\frac{1}{2}}\frac{\partial \eta}{\partial y}\Bigg\vert_{i,j-\frac{1}{2}}\right]-\frac{\lambda_{i,j} \Gamma_{i,j}  \nabla \eta \cdot \textbf{n}} {V_{ij}}  \\
           =-&\frac{1}{\Delta x}\left[\left(\zeta A \right)_{i+\frac{1}{2},j}u^{*}_{i+\frac{1}{2},j}-\left(\zeta A \right)_{i-\frac{1}{2},j}u^{*}_{i-\frac{1}{2},j} \right] \\
             -&\frac{1}{\Delta y}\left[\left(\zeta A \right)_{i,j+\frac{1}{2}}v^{*}_{i,j+\frac{1}{2}}-\left(\zeta A \right)_{i,j-\frac{1}{2}}v^{*}_{i,j-\frac{1}{2}}\right]+\frac{\lambda_{ij} \Gamma_{i,j} \text{v}_{\textbf{n}_{x}} }{V_{i,j}},\label{eq:disc_poiss}
\end{split}
\end{equation}
where this equation is easily expanded to three dimensions.

\subsection{Thermal Immersed Boundary Method}\label{sec:therm_IB}
\begin{figure}
\begin{subfigure}{.5\textwidth}
  \centering
  \includegraphics[width=1.0\linewidth, clip=true]{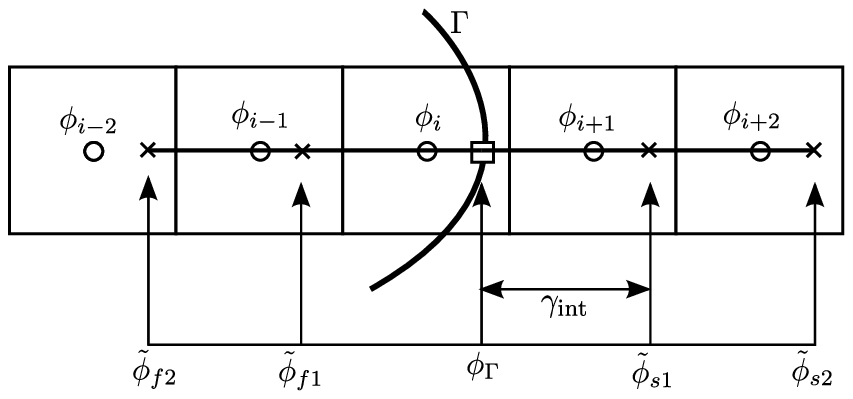}
  \caption{}
  \label{fig:cut_mid_stencil}
\end{subfigure}
\begin{subfigure}{.40\textwidth}
  \centering
  \includegraphics[width=0.9\linewidth, clip=true]{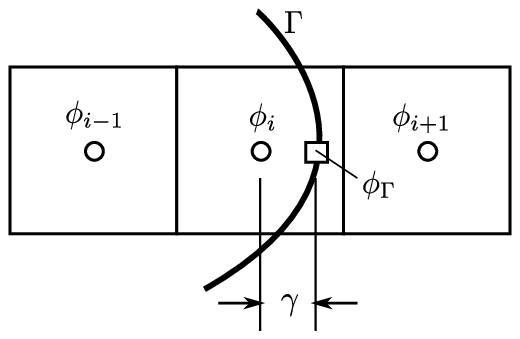}
  \caption{}
  \label{fig:cut_small_stencil}
\end{subfigure}
\caption{a.) One dimensional interpolation stencils used to find $\phi_{\Gamma}$. b.) One dimensional stencil used to apply the boundary condition from $\Gamma$ to $\phi_{i}$.}
\label{fig:CHTIB_stencils}
\end{figure}

The CHTIB derived form the conjugate heat transfer boundary conditions at the interface of two dissimilar substances without contact resistance.  Those boundary conditions are the continuity of the normal surface flux rates at $\Gamma$, shown in variable and discretized form, 
\begin{equation}
\kappa_{s}\frac{\partial \phi}{\partial \bn_{\Gamma s}} =  \kappa_{f}\frac{\partial \phi}{\partial \bn_{\Gamma f}}\rightarrow\kappa_{s}\frac{3\phi_{\Gamma s}-4\tilde{\phi}_{s1}+\tilde{\phi}_{s2}}{\partial \bn_{\Gamma s}}= \kappa_{f}\frac{-3\phi_{\Gamma f}+4\tilde{\phi}_{f1}-\tilde{\phi}_{f2}}{\partial \bn_{\Gamma f}}, \label{eq:cht_bc_1}
\end{equation}
respectively, and the continuity of temperature at $\Gamma$, 
\begin{equation}
 \phi_{\Gamma s} =\phi_{\Gamma f}=\phi_{\Gamma}, \label{eq:cht_bc_2}
\end{equation}
where the $\tilde{\left(\cdot\right)}$ terms are the interpolated equidistant (spaced at $\gamma_{\text{int}}$) from the interface on either side of the domain using second order interpolation scheme (see figure \ref{fig:cut_mid_stencil}) s, and $\kappa$ is the thermal conductivity.  The discretized forms of the CHT boundary conditions, given in equations \ref{eq:cht_bc_1} and \ref{eq:cht_bc_2}, are used to form an implicit, single boundary condition for any cut cell.  Substituting equation \ref{eq:cht_bc_2}, and the relationship $\bn_{\Gamma s}=-\bn_{\Gamma f}=\bn_{\Gamma}$, into equation \ref{eq:cht_bc_1} gives the new flux condition, 
\begin{equation}
\left .\kappa_{s}\frac{3\phi_{\Gamma}-4\tilde{\phi}_{s1}+\tilde{\phi}_{s2}}{\partial \bn_{\Gamma}}\right|_{s} = \left .\kappa_{f}\frac{-3\phi_{\Gamma}+4\tilde{\phi}_{f1}-\tilde{\phi}_{f2}}{\partial \bn_{\Gamma}}\right|_{f}. \label{eq:cht_bc_3}
\end{equation}
Equation \ref{eq:cht_bc_3} is solved for $\phi_{\Gamma}$ which is used to apply the CHT boundary condition to the cell containing $\phi_{i}$ from figure \ref{fig:cut_small_stencil}.

The CHTIB is implemented in the same manner as embedded FV boundary conditions.  Solving for $\phi_{\Gamma}$, using the interface distance, $\gamma$ from figure \ref{fig:cut_small_stencil}, the Crank-Nicholson, semi-implicit form, of the scalar transport equation, 
\begin{align}
& \partial_{t}\phi=\boldsymbol{\nabla} \cdot \left( \boldsymbol{\nabla}\alpha \phi \right) -\bnabla\cdot\left(\textbf{u}\phi\right) \rightarrow \nonumber \\
& a_{i}\phi^{n+1}_{i-1}+b_{i}\phi^{n+1}_{i}+c_{i}\phi^{n+1}_{i+1} =a_{i}\phi^{n}_{i-1}+b_{i}\phi^{n}_{i}+c_{i}\phi^{n}_{i+1}, \label{eq:CN_coeffs}
\end{align}
where $a$, $b$, and $c$ are the linear operator metrics from the FV discretization showing both the implicit, left hand side (LHS), and explicit, right hand side (RHS).  The embedded, one dimensional, boundary condition for figure \ref{fig:cut_small_stencil}, 
\begin{equation}
\phi^{n}_{i+1}=\phi^{n}_{\Gamma}\frac{\Delta x_{i}}{\gamma}+\phi^{n}_{i}\left(1-\frac{\Delta x_{i}}{\gamma}\right)\label{eq:embed_bound}
\end{equation}
given the implicit boundary conditions as,
\begin{equation}
a_{i}\phi^{n+1}_{i-1}+\phi^{n+1}_{i}\left[b_{i}-\left(1-\frac{\Delta x_{i}}{\gamma}\right) \right] \rightarrow c_{i}=0,\label{eq:embed_boundCNimp}
\end{equation}
and explicit as equation \ref{eq:embed_boundCNexp},
\begin{equation}
\boldsymbol{\nabla} \cdot \left( \boldsymbol{\nabla}\alpha \phi \right) -\bnabla\cdot\left(\textbf{u}\phi\right) \hspace{0.5cm}  \text{with} \hspace{0.5cm} \phi^{n}_{i+1}=\phi^{n}_{\Gamma}\frac{\Delta x_{i}}{\gamma}+\phi^{n}_{i}\left(1-\frac{\Delta x_{i}}{\gamma}\right), \label{eq:embed_boundCNexp}
\end{equation}
in the flux terms.  By replacing $\gamma$ in the boundary condition interpolation, equation \ref{eq:embed_bound} it can be noted that equations \ref{eq:embed_boundCNimp} and \ref{eq:embed_boundCNexp} reduced to standard, second order embedded boundary conditions using ghost nodes.  It should also be noted that the values of either right or left boundary conditions, $\phi_{i+1}$ or $\phi_{i-1}$ are not replaced in the solution, only their flux terms are affected. 

There are some stability issues with the final form of equations \ref{eq:embed_boundCNimp} and \ref{eq:embed_boundCNexp}. The first, is $\lim_{\gamma \rightarrow 0}\frac{\Delta}{\gamma}=\infty$ causes solution instability when $\gamma/\Delta x\le 0.05$.  By keeping the equations \ref{eq:embed_boundCNimp} and \ref{eq:embed_boundCNexp} in their given form for $\gamma/\Delta x\ge 0.05$, and replacing the interpolation of the embedded boundary condition in equation \ref{eq:embed_bound} with $\phi_{i}=\phi_{\Gamma}$, removes this stability constraint by adding a simple Dirichlet boundary condition to the point collocated on $\Gamma$.  

The second stability concern is how to treat $\phi^{n}_{\Gamma}$ and $\phi^{n+1}_{\Gamma}$.  It is clear from the form of equation \ref{eq:embed_boundCNimp} that it has already been taken that $\phi^{n}_{\Gamma}= \phi^{n+1}_{\Gamma}$.  This was found to be consistent with analytical solutions, and using a $dt$ of the order used for the smallest scales of $\phi$, will give second order results (see section \ref{sec:TIBval}).  

\begin{figure}
\begin{subfigure}{.45\textwidth}
  \centering
  \includegraphics[width=1.0\linewidth, clip=true]{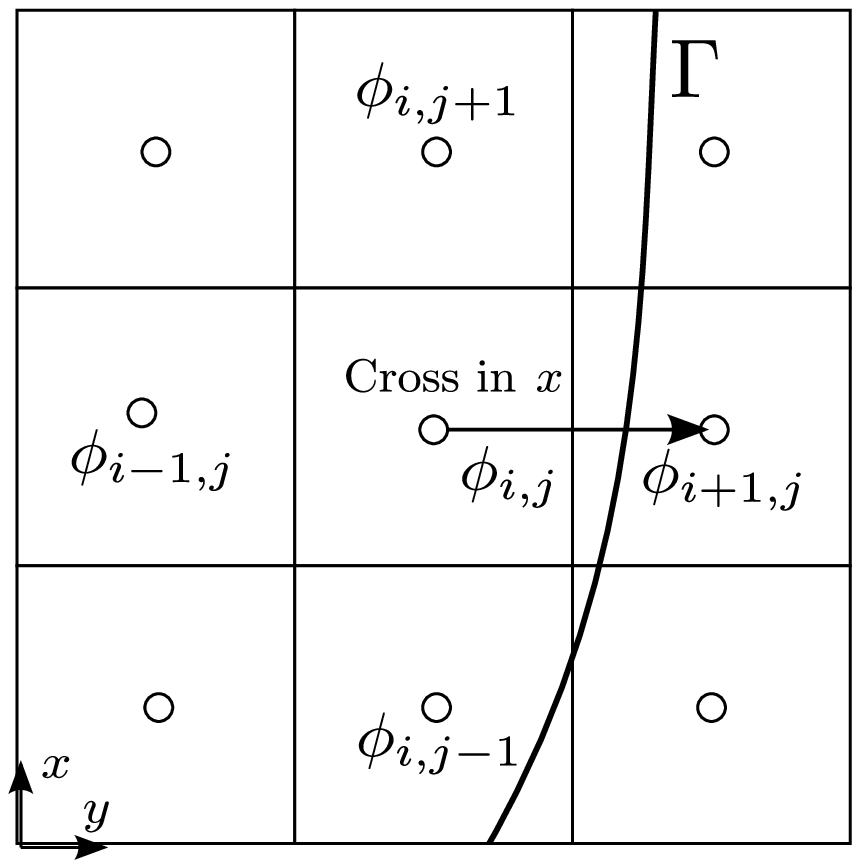}
  \caption{}
  \label{fig:cut_X}
\end{subfigure}
\begin{subfigure}{.45\textwidth}
  \centering
  \includegraphics[width=1.0\linewidth, clip=true]{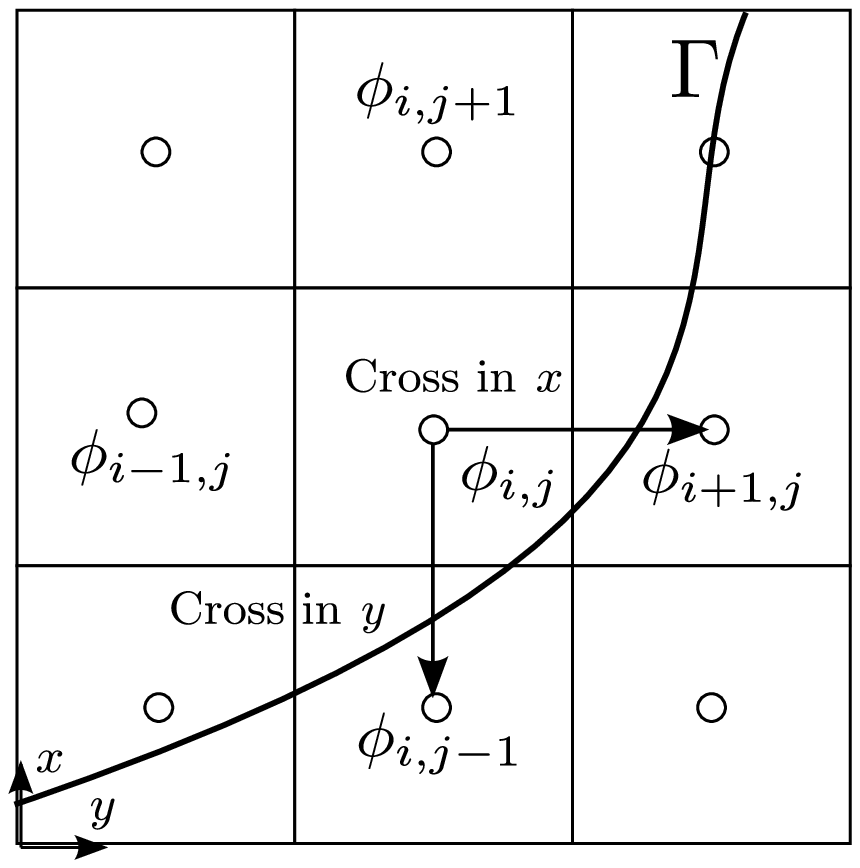}
  \caption{}
  \label{fig:cut_XY}
\end{subfigure}
\caption{a.) Two dimensional schematic showing how $\Gamma$ would forced the CHT boundary conditions to be applied at $\phi_{i+t,j}$. b.) Two dimensional schematic showing how $\Gamma$ would forced the CHT boundary conditions to be applied at $\phi_{i+t,j}$ and  $\phi_{i,j-1}$.}
\label{fig:CHTIB_examples}
\end{figure}

The third, and final stability constraint has to do with implementation in $N$ dimensions.  When a FV cell is cut only the fluxes in the direction that will cross $\Gamma$ is used to determine the form of the evolution equation for that cell.  For example, in the cells in shown in figure \ref{fig:CHTIB_examples}, in figure \ref{fig:cut_X} the CHT boundary condition would be applied for the $\phi_{i+1,j}$ and the fluxes in the $y$ would be set equal at each face of the FV cell at $\{i,j-1\}$ and $\{i,j+1\}$.  In figure \ref{fig:cut_XY} the boundary lies across both that FV cell faces connect $\phi_{i,j}$ to $\phi_{i+1,j}$ and $\phi_{i,j-1}$.  The CHT boundary conditions would then be applied at both of those points.  Canceling out the fluxes, near $\Gamma$, that do not apply directly to a CHT boundary for that cell is justified by the physical assumptions that: near the interface the fluxes not orthogonal to $\ \textbf{n}_{\Gamma}$ will not conduct heat across that interface, 
\begin{equation}
\textbf{F}\times \textbf{n}_{\Gamma} =0 \hspace{0.25cm} \text{if} \hspace{0.25cm} \textbf{F}\parallel  \textbf{n}_{\Gamma}, \label{eq:cross_prod}
\end{equation}
constraining the near interface physics to CHT.  Second, iso-therms near $\Gamma$ will be parallel to it giving, 
\begin{equation}
\bnabla \phi\parallel \bn_{\Gamma}<<\bnabla \phi\perp \textbf{n}_{\Gamma}. \label{eq:grad_big_small}
\end{equation}
Thirdly, the simple numerical constraint that there is no sub-grid scale model being used for clipped cells, other than the CHT one laid previously.

Essentially the CHTIB decouples the fluid and solid domains by leveraging the the FVM.  The resulting form transfers information with high order interpolation methods which pass through the CHT boundary conditions.  In this way, either domain can only indirectly affect the other, and only in a way that is prescribed in equations \ref{eq:cht_bc_1}-\ref{eq:cht_bc_2}.

\section{Thermal IB Results and Validation }\label{sec:TIBval}

\subsection{Co-Annular Validation} \label{sec:coannval}
\begin{figure}
\begin{subfigure}{.45\textwidth}
  \centering
  \includegraphics[width=1.0\linewidth, clip=true]{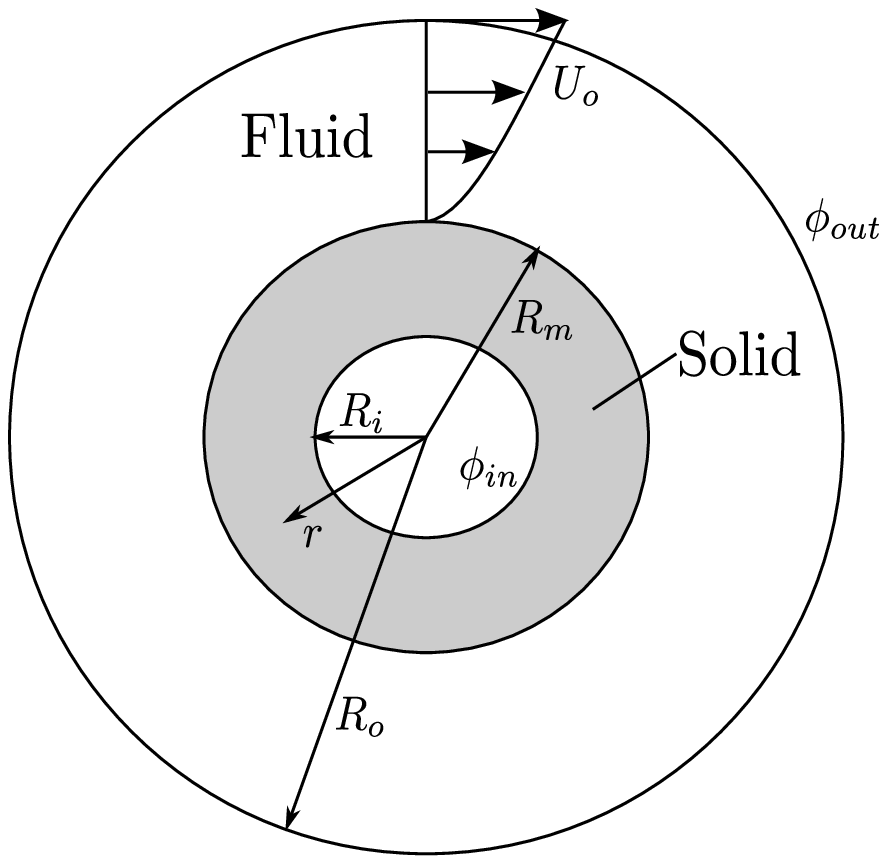}
  \caption{}
  \vspace{0.1cm}
  \label{fig:val_sch}
\end{subfigure}
\hfill
\begin{subfigure}{.45\textwidth}
  \centering
  \includegraphics[width=1.0\linewidth, clip=true]{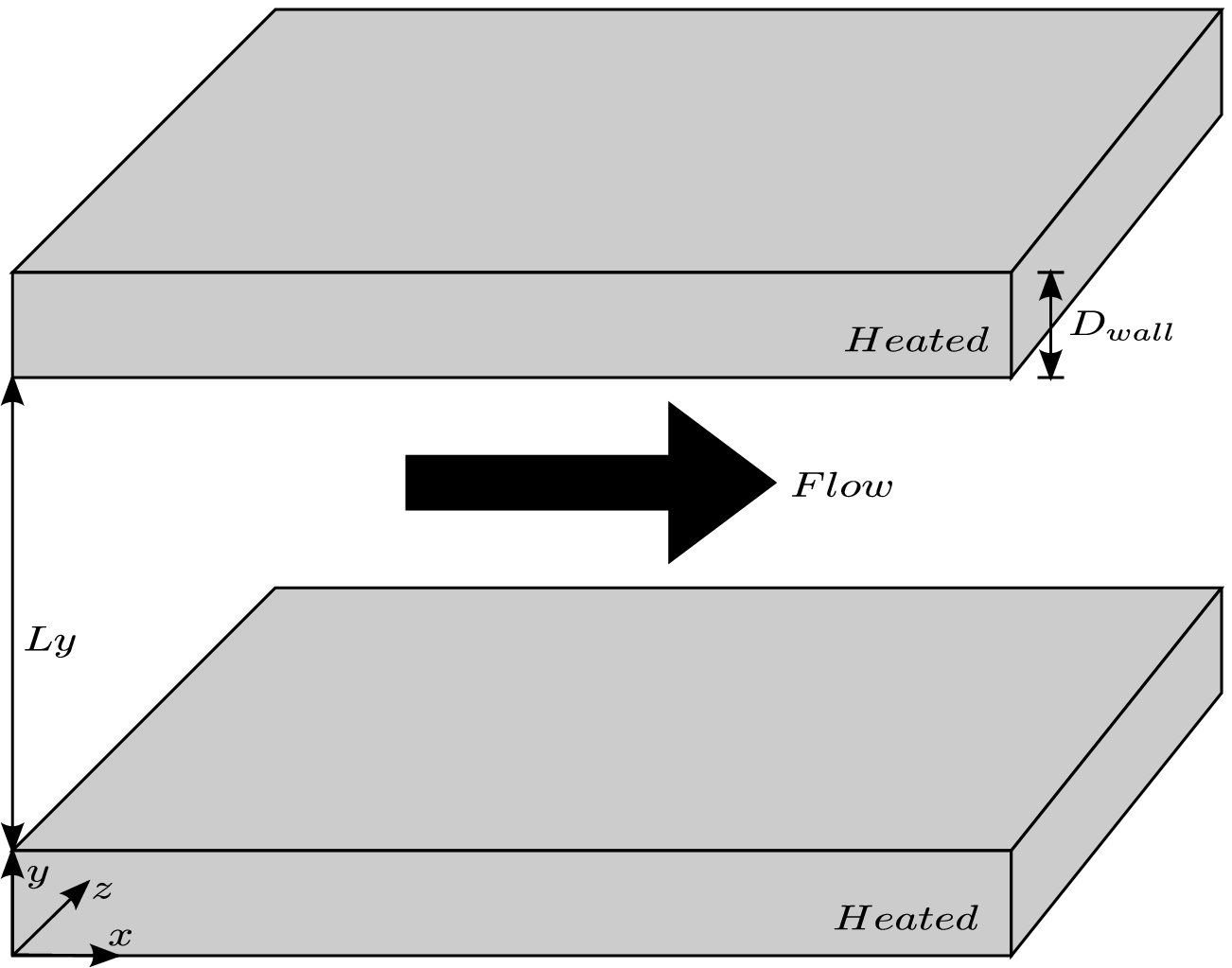}
  \caption{}
\vspace{0.1cm}
  \label{fig:abl_stencil}
\end{subfigure}
\caption{a.) Schematic of the co-annular validation study domain.  b.) Schematic of the heated channel validation study domain periodic in $x$ and $z$ directions.}
\label{fig:val_chan}
\label{fig:CHTIB_schem}
\end{figure}

Validation was done with one of the few CHT analytical solution, using the normalization of equations \ref{eq:navier_stokes_dim}-\ref{eq:continuity}, for incompressible flow, given as, 
\begin{equation}
\partial_{t^{*}}\left(\textbf{u}^{*}\right) + \bnabla^{*} \cdot \left(\textbf{u}^{*}  \textbf{u}^{*} \right)= -\bnabla^{*} \cdot \textbf{I}P^{*}+ \frac{1}{Re}\bnabla^{*} \cdot \left(\bnabla^{*} \textbf{u}^{*}\right), \label{eq:Norm1navier_stokes_dim}
\end{equation}
\begin{equation}
\partial_{t^{*}}\left(\textbf{u}^{*}\right) +\bnabla^{*} \cdot \left( \textbf{u}^{*}\right)=0. \label{eq:Norm1continuity}
\end{equation}
The equations for heat transfer, equations \ref{eq:energy}-\ref{eq:energy_solid}, are normalized as, 
\begin{equation}
\partial_{t^{*}}\phi^{*}+\bnabla^{*} \cdot \left(\textbf{u}^{*}\phi^{*}\right) = \frac{1}{RePr}\bnabla^{*} \cdot \left( \bnabla^{*} \phi^{*}\right),  \label{eq:Norm1energy}
\end{equation}
\begin{equation}
\partial_{t}\phi^{*} = \frac{1}{RePr}\frac{\alpha_{s}}{\alpha_{f}}\bnabla^{*} \cdot \left( \bnabla^{*} \phi^{*}\right),  \label{eq:Norm1energy_solid}
\end{equation}
where $Re$ is the Reynolds number and $Pr$ is the Prandtl number.  The dimensionless groups used in the normalizations in equations \ref{eq:Norm1navier_stokes_dim}-\ref{eq:Norm1continuity} and \ref{eq:Norm1energy}-\ref{eq:Norm1energy_solid} are, 
\begin{equation}
\begin{split}
\textbf{u}^{*}=\frac{\textbf{u}}{U_{\infty}}  \label{eq:annval_Re_group}
\end{split}
\hspace{0.5cm}
\begin{split}
 \textbf{x}^{*}=\frac{\textbf{x}}{D_{o}}
\end{split}
\hspace{0.5cm}
\begin{split}
 t^{*}=t\frac{U_{\infty}}{D_{o}} 
\end{split}
\hspace{0.5cm}
\begin{split}
P^{*}=\frac{P}{\rho U_{\infty}^{2}}
\end{split}
\hspace{0.5cm}
\begin{split}
Re=\frac{U_{\infty}D_{o}}{\nu},
\end{split}
\end{equation}
where $D_{o}=2R_{o}$, and $U_{\infty}=u_{r}\left(R_{o} \right)$, defined in figure \ref{fig:CHTIB_schem} and equations \ref{eq:mal_T} and \ref{eq:anna_T}.

\begin{figure}
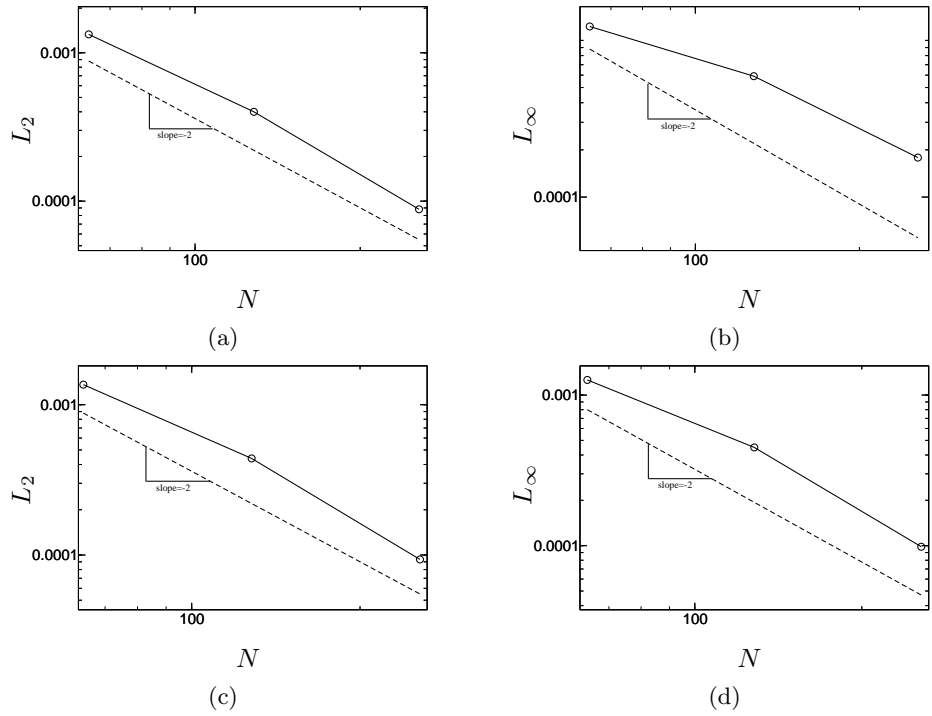

\begin{subfigure}{.45\textwidth}
  \centering
    \psfrag{xaxis}{$N$}
    \psfrag{yaxis}{$L_{2}$}
  \includegraphics[width=1.0\linewidth, clip=true]{figures/L2_temp.eps}
  \caption{}
  \vspace{0.1cm}
  \label{fig:L2_sk}
\end{subfigure}
\hfill
\begin{subfigure}{.45\textwidth}
  \centering
      \psfrag{xaxis}{$N$}
    \psfrag{yaxis}{$L_{\infty}$}
  \includegraphics[width=1.0\linewidth, clip=true]{figures/L2_inf.eps}
  \caption{}
\vspace{0.1cm}
  \label{fig:Linf_sk}
\end{subfigure}
\begin{subfigure}{.45\textwidth}
  \centering
    \psfrag{xaxis}{$N$}
    \psfrag{yaxis}{$L_{2}$}
  \includegraphics[width=1.0\linewidth, clip=true]{figures/L2_temp_BK.eps}
  \caption{}
  \label{fig:L2_bk}
\end{subfigure}
\hfill
\begin{subfigure}{.45\textwidth}
  \centering
        \psfrag{xaxis}{$N$}
    \psfrag{yaxis}{$L_{\infty}$}
  \includegraphics[width=1.0\linewidth, clip=true]{figures/L2_error_BK.eps}
  \caption{}
  \label{fig:Ling_bk}
\end{subfigure}
\caption{a.) The $L_{2}$ error norm for temperature and b.) the $L_{\infty}$ error norm for temperature both for  $\kappa_{s}/\kappa_{f}=9$.  c.) The $L_{2}$ error norm for temperature, and d.) The $L_{\infty}$ error norm for temperature, both for  $\kappa_{s}/\kappa_{f}=900$.  All four plots show second order convergence of error, $ - - -$ is a line with LOG-LOG slope of -2 and $\circ-\circ-\circ$ are the error norms, they are both plotted against the number of nodes, $N=Nx=Ny$ in the simulation.}
\label{fig:error_norms}
\end{figure}

\begin{figure}
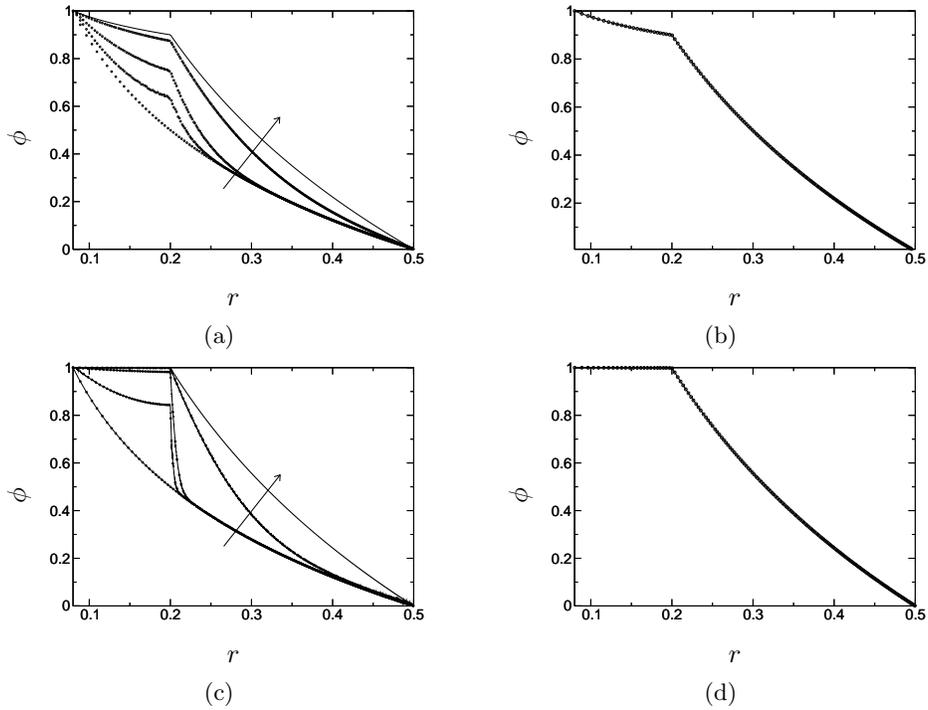

\begin{subfigure}{.45\textwidth}
  \centering
    \psfrag{xaxis}{$r$}
    \psfrag{yaxis}{$\phi$}
  \includegraphics[width=1.0\linewidth, clip=true]{figures/annval_timeseries_256_annval.eps}
  \caption{}
  \vspace{0.1cm}
  \label{fig:timeseries_256_annval}
\end{subfigure}
\hfill
\begin{subfigure}{.45\textwidth}
  \centering
      \psfrag{xaxis}{$r$}
    \psfrag{yaxis}{$\phi$}
  \includegraphics[width=1.0\linewidth, clip=true]{figures/annval_256_final.eps}
  \caption{}
\vspace{0.1cm}
  \label{fig:annval_256_final}
\end{subfigure}
\begin{subfigure}{.45\textwidth}
  \centering
      \psfrag{xaxis}{$r$}
    \psfrag{yaxis}{$\phi$}
  \includegraphics[width=1.0\linewidth, clip=true]{figures/256_timeseries_BK.eps}
  \caption{}
  \label{fig:timeseries_BK}
\end{subfigure}
\hfill
\begin{subfigure}{.45\textwidth}
  \centering
      \psfrag{xaxis}{$r$}
    \psfrag{yaxis}{$\phi$}
  \includegraphics[width=1.0\linewidth, clip=true]{figures/256_final_BK.eps}
  \caption{}
  \label{fig:final_BK}
\end{subfigure}
\caption{a.) The time series for the temperature profile plotted with the analytical solution, b.) and the final temperature profile plotted against  the analytical solution for$\kappa_{s}/\kappa_{f}=9$.  c.) The time series for the temperature profile plotted with the analytical solution, d.) and the final temperature profile plotted against  the analytical solution for$\kappa_{s}/\kappa_{f}=900$, lines have been added to the time series to help distinguish each time step.  $-----$ is the analytical solution $\circ-\circ-\circ$ are the temperature profiles increasing with time in direction of the arrow.  All plots shown are from the $N_{x}=256$, $N_{y}=256$ simulations with every two hundredth point removed from the plot to make it easier to see individual points.}
\label{fig:annval_results}
\end{figure}

The velocity and temperature fields for two co-annular cylinders (see figure \ref{fig:val_sch}) are given by, 
\begin{equation}
u_{\theta} = 0, \hspace{0.25cm} \text{and} \hspace{0.25cm}  u_{r}\left(r\right) = 
\begin{dcases}
 0 &\text{for} \hspace{0.2cm} R_{i}<r<R_{m} \\
 -\frac{R_{o}R^{2}_{m}U_{\infty}}{R^{2}_{o}-R^{2}_{m}}\frac{1}{r}+\frac{R_{o}U_{\infty}}{R^{2}_{o}-R^{2}_{m}}r &\text{for} \hspace{0.2cm} R_{m}<r<R_{o}
\end{dcases}\label{eq:mal_T}
\end{equation}
in cylindrical coordinates, and, 
\begin{equation}
\phi_{r}\left(r\right) = 
\begin{dcases}
 \phi_{in}+\frac{\phi_{out}-\phi_{in}}{\text{log}_{10}\left(\frac{R_{m}}{R_{i}}\right)+\left(\frac{\kappa_{s}}{\kappa_{f}}\right) \text{log}_{10}\left(\frac{R_{o}}{R_{m}}\right)}\text{log}_{10}\left(\frac{r}{R_{i}}\right) &\text{for} \hspace{0.2cm} R_{i}<r<R_{m} \\
 \phi_{out}-\frac{\phi_{out}-\phi_{in}}{\left(\frac{\kappa_{f}}{\kappa_{s}}\right) \text{log}_{10}\left(\frac{R_{m}}{R_{i}}\right)+\text{log}_{10}\left(\frac{R_{o}}{R_{m}}\right)}\text{log}_{10}\left(\frac{R_{o}}{r}\right)&\text{for} \hspace{0.2cm} R_{m}<r<R_{o}
\end{dcases}\label{eq:anna_T}
\end{equation}
respectively, where the dimensions are given in figure \ref{fig:val_sch} and $Re=50$.
The figures \ref{fig:error_norms} a,b,c, and d, show the $L_{2}=\sum\left({\phi_{exact}-\phi}\right)^2/\sum{\phi_{exact}^2}$ and $L_{\infty}=\text{max}|\phi_{exact}-\phi |$ norms after reaching the convergence criteria of $L_{2}^{n+1}-L_{2}^{n}|\le1\cdot 10^{-9}$ ($\approx 20000$ iterations at $dt=1\cdot 10^{-4}$), and a steady state.
The time series show in figure \ref{fig:annval_results} gives good indication that even in transience the CHT equations are still holding the time evolution of the temperature profiles to the sharp boundary at the discontinuity of $\kappa$.   Even with $\kappa_{s}>>\kappa_{f}$, as shown in figure \ref{fig:timeseries_BK} the initially discontinuity of the time series is nearly square and the solution remains stable, even in the points close to the interface.  During the development of the CHT algorithm defined in section \ref{sec:therm_IB} experiments were initially run purely explicitly and it was noted that as the ratio $\kappa_{s}/\kappa_{f}$ was increased so was the temporal stiffness of the time evolution equation making large $\kappa$ rations slow and unwieldy.  The edition of the CHT boundary conditions to an implicit solver seems to have mitigated the issues of small $dt$ for large $\kappa_{s}/\kappa_{f}$.   

For both the plots in figures \ref{fig:annval_256_final} and \ref{fig:final_BK} the final steady state lays directly over the analytical solution given in equation \ref{eq:anna_T}, showing very little deviation at any point including interfacial points.  Two different were used to find the final steady state.  Initiating the temperature fields to equation \ref{eq:anna_T} and waiting until the error norms converged and letting the fields pass through transience to the steady state solution.  Both simulation types produced indistinguishable error plots (figure \ref{fig:error_norms}), that were second order convergent for $L_{\infty}$, and $L_{2}$.

\subsection{Turbulent Heated Channel Validation} \label{sec:chanval}

The second numerical experiment for validation of the CHTIB is  a turbulent heated channel originally put forward in \cite{Kasagi1989} with the addition of CHT added in \cite{Tiselj2001}.  The geometry of the heated channel domain follows the simple geometry in figure \ref{fig:val_chan}.  The normalization in previous works, \cite{Kasagi1992}, \cite{Kawamura1998}, and \cite{Tiselj2001}, was not applicable to the momentum and pressure solver used in this work so the general equations, here, were normalized as put forward in equations \ref{eq:Norm1navier_stokes_dim}-\ref{eq:Norm1energy_solid}.  With the exception that the normalized groups are now based on the channel half width $L_{y}/2$ and the skin friction velocity at the wall, $u_{\tau}=\sqrt{\mu\left( du/dy\right)_{wall}}$,
\begin{equation}
\begin{split}
\textbf{u}^{*}=\frac{\textbf{u}}{u_{\tau}}  \label{eq:chanRe_group}
\end{split}
\hspace{0.5cm}
\begin{split}
 \textbf{x}^{*}=2\frac{\textbf{x}}{L_{y}}
\end{split}
\hspace{0.5cm}
\begin{split}
 t^{*}=2t\frac{u_{\tau}}{L_{y}} 
\end{split}
\hspace{0.5cm}
\begin{split}
P^{*}=\frac{P}{\rho u_{\tau}^{2}}
\end{split}
\hspace{0.5cm}
\begin{split}
Re_{u_{\tau}}=2\frac{u_{\tau}L_{y}}{\nu},
\end{split}
\end{equation}
\begin{figure}
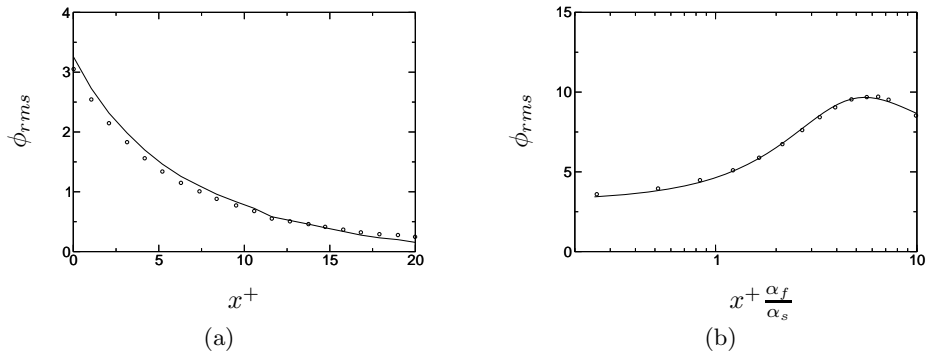

\begin{subfigure}{.45\textwidth}
  \centering
      \psfrag{xaxis}{$x^{+}$}
    \psfrag{yaxis}{$\phi_{rms}$}
  \includegraphics[width=1.0\linewidth, clip=true]{wall_plot.eps}
  \caption{}
  \label{fig:chan_wall}
\end{subfigure}
\hfill
\begin{subfigure}{.45\textwidth}
  \centering
      \psfrag{xaxis}{$x^{+}\frac{\alpha_{f}}{\alpha_{s}}$}
    \psfrag{yaxis}{$\phi_{rms}$}
  \includegraphics[width=1.0\linewidth, clip=true]{core_plot.eps}
  \caption{}
  \label{fig:chan_core}
\end{subfigure}
\caption{a.) The root-mean-square $\left( rms\right)$ of the fluctuating temperature inside the heated wall b.) The  $\left( rms\right)$ of the fluctuating temperature inside the core flow.  For both a.) and b.) the wall units are those taken from \cite{Tiselj2001}, as well as the scaling of the temperature field inside the wall and the core flow. Current results, \pline and results from \cite{Tiselj2001}, $\circ-\circ-\circ$.}
\label{fig:channel_results}
\end{figure}d
and the addition of a source term to equation \ref{eq:Norm1energy_solid}, 
\begin{equation}
\partial_{t}\phi^{*} = \frac{1}{RePr}\frac{\alpha_{s}}{\alpha_{f}}\bnabla^{*} \cdot \left( \bnabla^{*} \phi^{*}\right)+\frac{\alpha_{f}}{\alpha_{s}}\frac{1}{Re_{u_{\tau}}D_{\text{wall}}}.  \label{eq:Norm2energy_solid}
\end{equation}
where $\alpha_{s}\alpha_{f}=4$, and $Pr=7$ and the term to the to the left on the right hand side of equation \ref{eq:Norm2energy_solid} is the heating source term. The boundary of the channel domain at $y=\pm \left(\frac{L_{y}}{2}+D_{\text{wall}}\right)$ is kept adiabatic.  At $\Gamma$ the ensemble average of temperature, in time and space, is $\overline{\phi}=0$ with the no slip condition for velocity and a constant pressure gradient.  The mesh resolution and stretching are, $Lx=64$, $Ly=176$, $Lz=64$, and $\delta y^{+}=0.029\leftrightarrow2.95$.  The channel is simulated until the momentum fields reach a statistical steady state where the turbulent production is balanced by the dissipative terms \cite{popebook}. 


in The results from \cite{Tiselj2001} are scaled with the normalization used in equation \ref{eq:chanRe_group} and compared with the results from this work in figure \ref{fig:channel_results}.

\section{Conclusion}\label{sec:conclusion}
Analysis of the CHTIB algorithm shows second order accuracy with a sharp and distinct definition between the boundaries of the fluid and solid domain.  With implicit implementation a large rage of $\kappa_{s}/\kappa_{f}$ can be simulated without stability or accuracy concerns.   Testing to larger Reynolds numbers with 3D chaotic flow also shows that it is correlated with previous studies giving accurate results temperature field results for large $Pr$.  The momentum IB also shows that there was little detectable effect on the subsequent scalar field through the convective terms. The consequence is global second order error for the scalar fields and the first order momentum statistics

The addition of a level set field to define $\Gamma$ is a simple way to store geometric data for the cut-cell IB geometries used by both the momentum IB and the CHTIB, with a low computational and front end coding costs.  With both the CHTIB and momentum IB activated channel simulations for equal numbers of grid points show a $7-10\%$ decrease in computational efficiency, overall, but consistency with core scale up was little effected. 

\section{Acknowledgments}\label{sec:acknowledgments}
Our funding through the grants NSF CBET-0967857 and NASA NNX11AM07A. Professor Olivier Desjardins for the use of his NGA flow solver, and Dr. Michael Wright and Dr. Nagi Mansour for all their help at NASA AMES. 

\pagebreak

\section*{References}

\bibliography{CHT_arxiv}

\end{document}